\documentclass[useAMS,usenatbib]{mn2e}
\usepackage{epsfig}
\usepackage{amsmath, amssymb,bm}

\title[Propagation of the Gravo-Magneto Disc Instability]{Propagation
  of the Gravo-Magneto Disc Instability}

\author[R. G. Martin \& S. H. Lubow]{Rebecca
  G. Martin$^1$\thanks{E-mail: rebecca.martin@jila.colorado.edu} and
  Stephen H. Lubow$^2$\\ $^1$NASA Sagan Fellow, JILA, University of
  Colorado, Boulder, CO 80309, USA\\ $^2$Space Telescope Science
  Institute, 3700 San Martin Drive, Baltimore, MD 21218, USA \\ }

\begin{document}

\date{}

\pagerange{\pageref{firstpage}--\pageref{lastpage}} 
\pubyear{2013}
\maketitle

\label{firstpage}

\begin{abstract}
Discs that contain dead zones are subject to the Gravo-Magneto (GM)
instability that arises when the turbulence shifts from gravitational
to magnetic. We have previously described this instability through a
local analysis at some radius in the disc in terms of a limit cycle.
A disc may be locally unstable over a radial interval.  In this paper,
we consider how the local instability model can describe global disc
outbursts.  The outburst is triggered near the middle of the range of
locally unstable radii.  The sudden increase in turbulence within high
surface density material causes a snow plough of density that
propagates both inwards and outwards.  All radii inside of the trigger
radius become unstable, as well as locally unstable radii outside the
trigger radius. In addition, a locally stable region outside of the
trigger radius may also become unstable as the gravitational
instability is enhanced by the snow plough. For the circumstellar disc
model we consider, we find that a quarter of the disc mass is accreted
on to the central object during the outburst.  The radius out to which
the disc is globally unstable is twice that for which it is locally
unstable.
\end{abstract}


\section{Introduction}

Accretion discs are ubiquitous in the Universe, they form on all
scales from planetary, to stellar, to galactic. They transport angular
momentum outwards allowing material to spiral inwards through the disc
\cite[e.g.][]{pringle81}. If the disc is fully ionised, the magneto
rotational instability (MRI) drives turbulence and thus angular
momentum transport \citep{balbus91}. However, below a critical level
of ionisation, a dead zone may form at the mid-plane where the MRI
does not operate \citep{gammie96, gammie98, turner08}. The inner parts
of the disc are hot enough to be thermally ionised. However, further
out, only the surface layers of the disc may be ionised by external
sources such as cosmic rays or X-rays from the central star
\citep{glassgold04}.  The restricted flow through the disc causes a
build up of material in the dead zone. If there is sufficient
accretion infall, some turbulence in the dead zone may be driven by
self-gravity when the layer becomes massive enough
\citep{paczynski78,lodato04}. However, if it does not transport
angular momentum fast enough, the gravo-magneto (GM) disc instability
can result. This occurs when the temperature in the dead zone reaches
the critical required for the MRI to be triggered and an accretion
outburst ensues \citep{armitage01,zhu09,zhu10a,zhu10b,martin11}.

The GM disc instability can be explained as transitions between steady
state disc solutions plotted on a state diagram showing the accretion
rate through the disc against the surface density for a fixed radius
\citep{martin11}. There are three steady state disc solutions, two
fully MRI turbulent solutions (one that is thermally ionised and one
that is fully ionised by external sources such as cosmic rays or
X-rays) and the GM solution. The GM solution consists of a self
gravitating, MRI inactive mid-plane region with MRI active surface
layers such that the flow through the disc is at the steady rate. For
a given radius, there may be a range of accretion rates for which no
steady solution exists. If the accretion rate on to the disc lies in
this range then the disc will be locally GM unstable. The disc
transitions between the steady thermal MRI and GM solutions cause
repeating outbursts over time. This is similar to the ``S-curve'' used
to explain dwarf nova outbursts \citep{bath82, faulkner83}.

For a range of radii in the disc, there exists a local limit cycle. In
this work, we extend the model of \cite{martin11} to consider how the
local limit cycles at different radii are coordinated and thus how the
global disc evolution operates.  We examine the radius at which the
outburst is triggered and how far the outburst propagates through the
disc. We find that the radii are connected through a snow plough
effect where the dense material is pushed both inwards and outwards
during the outburst.  Previous works have considered the propagation
of the outburst \citep[e.g.][]{armitage01,zhu10b}, but not within the
framework of the limit cycle.

The GM disc instability may occur in accretion discs on many
scales. It is thought to describe FU Orionis outbursts in discs around
young stellar objects
\citep[e.g.][]{armitage01,zhu10b,martin12a,martin12b,stramatellos12}.
It is also likely to occur in circumplanetary discs that form around
massive planets as they form within the circumstellar disc
\citep{martin11b,lubow12a,lubow12b}. In this work we restrict our
analysis to that of a circumstellar disc, but note that the underlying
mechanism is the same for all scales.

\section{Local Disc Instability}

The layered disc model we apply is based on \cite{armitage01} and
further developed in \cite{zhu10b} and \cite{martin11}. Following
\cite{martin11}, we first solve the steady state layered accretion
disc equation
\begin{equation}
\dot M =3 \pi ( \nu_{\rm m} \Sigma_{\rm m} + \nu_{\rm g} \Sigma_{\rm g})
\end{equation}
with the steady energy equation
\begin{equation}
\sigma T^4= \frac{9}{8} \Omega^2(\tau_{\rm m} \nu_{\rm m} \Sigma_{\rm m} + \tau \nu_{\rm g} \Sigma_{\rm g}),
\end{equation}
where $\dot M$ is the steady accretion rate through the disc, $T$ is
the midplane temperature, $\Sigma_{\rm g}$ is the surface density in
the dead zone layer, $\Sigma_{\rm m}$ is the MRI active surface
density, the total surface density is $\Sigma=\Sigma_{\rm
  m}+\Sigma_{\rm g}$, $\Omega$ is the Keplerian angular velocity,
$\tau_{\rm m}$ is the optical depth of the magnetic layer and $\tau$
is the optical depth of the whole disc. The viscosity due to the MRI
is
\begin{equation}
\nu_{\rm m}=\alpha_{\rm m}\frac{c_{\rm m}^2}{\Omega},
\end{equation}
where $c_{\rm m}$ is the sound speed in the active surface layers and
$\alpha_{\rm m}$ is the \cite{shakura73} $\alpha$ parameter in the
layer. The disc is self-gravitating if the \cite{toomre64} parameter
is less than the critical, $Q<Q_{\rm crit}=2$, and in this case
$\nu_{\rm g}$ depends on $Q$. The viscosity due to self-gravity is
\begin{equation}
\nu_{\rm g}=\alpha_{\rm g}\frac{c_{\rm g}^2}{\Omega},
\end{equation}
where $c_{\rm g}$ is the sound speed in the mid-plane layer and
\begin{equation}
\alpha_{\rm g}=\alpha_{\rm m} \left[\left(\frac{Q_{\rm crit}}{Q}\right)^2-1\right]
\end{equation}
for $Q<Q_{\rm crit}$ and zero otherwise \citep{lin87,lin90}.  We find
that the exact form of $\alpha_{\rm g}$ does not affect the disc
evolution provided it is a steeply declining function of $Q$
\citep[see also][and the Discussion section]{zhu10b}.

In this work we consider a constant infall accretion rate, $\dot
M_{\rm infall}$, on to the disc. For a steady disc solution, the
infall accretion rate is equal to the accretion rate through the disc,
$\dot M$, at all radii. However, if at some radius there is no steady
solution, then the disc is unstable to the GM disc instability.

As described in the Introduction, there are three different steady
solutions, two MRI active solutions and the GM solution. The fully MRI
active solutions are found by setting $\nu_{\rm g}=0$. They exist in
two regions of the disc, first in the inner parts where the mid-plane
temperature is sufficiently high for thermal ionisation ($T>T_{\rm
  crit}$, where $T_{\rm crit}$ is the temperature above which the MRI
operates) and secondly in the outer parts where the surface density is
sufficiently small that cosmic rays or X-rays penetrate the whole disc
($\Sigma<\Sigma_{\rm crit}$, where $\Sigma_{\rm crit}$ is the critical
surface density that is ionised through external sources). The third
steady solution is the GM solution, where there is gravitational
turbulence near the mid-plane and MRI turbulence in disc surface
layers.  This solution occurs where the mid-plane temperature is too
low for the MRI ($T<T_{\rm crit}$) and the surface density is greater
than the critical that can be ionised by external sources
($\Sigma>\Sigma_{\rm crit}$), and large enough for the disc to be
self-gravitating. The steady state solutions at a given radius cover a
range of accretion rates. In general, there are cases where there is a
gap in accretion rates for steady state solutions. Accretion rates
that fall in that gap at some radius are locally unstable
\citep{martin11}. For a fixed accretion rate on to the disc, there is
generally a range of radii where that rate falls within the local gaps
and thus a range of radii where the disc can be locally unstable.

\begin{figure*}
\includegraphics[width=7cm]{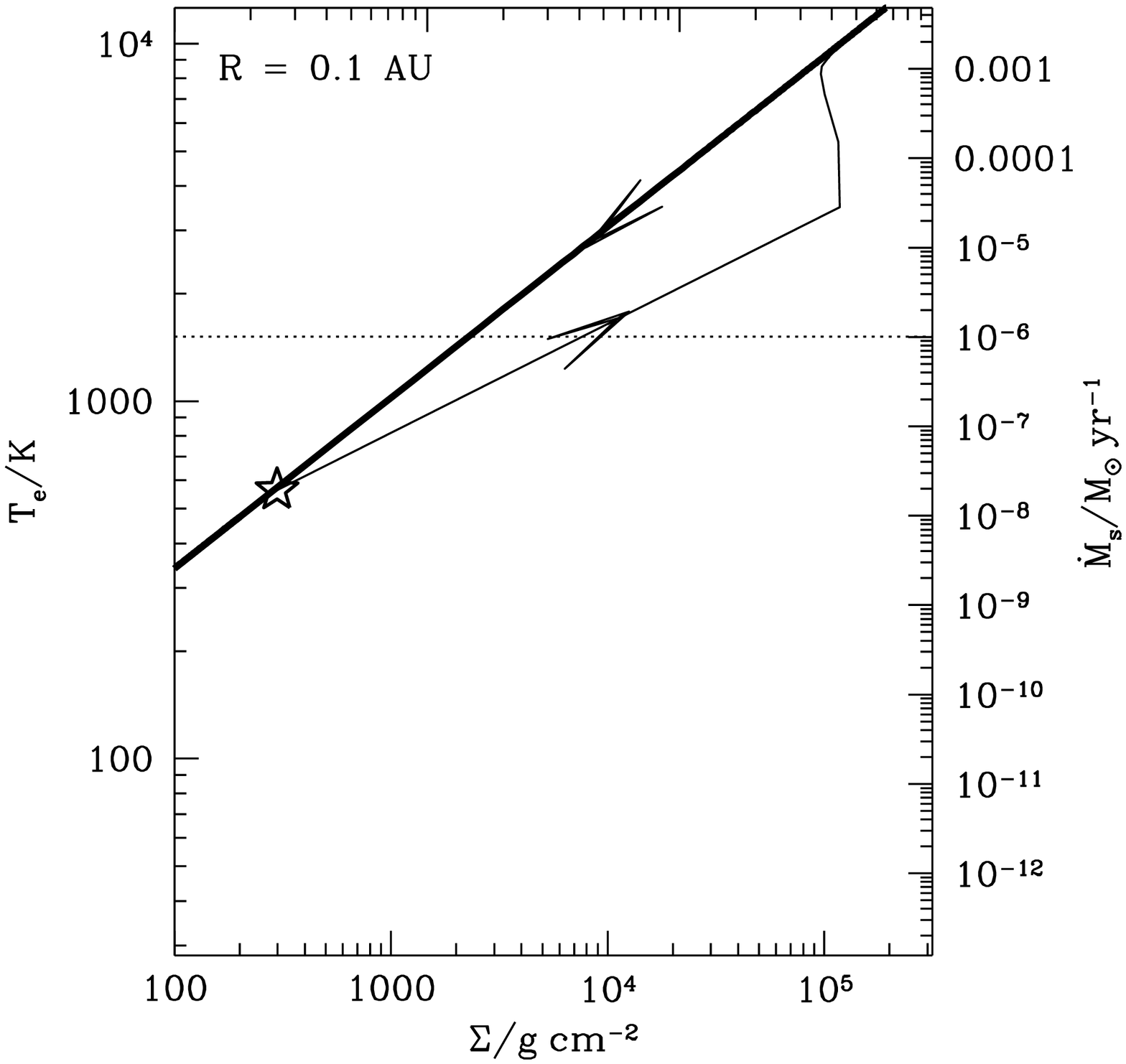}
\includegraphics[width=7cm]{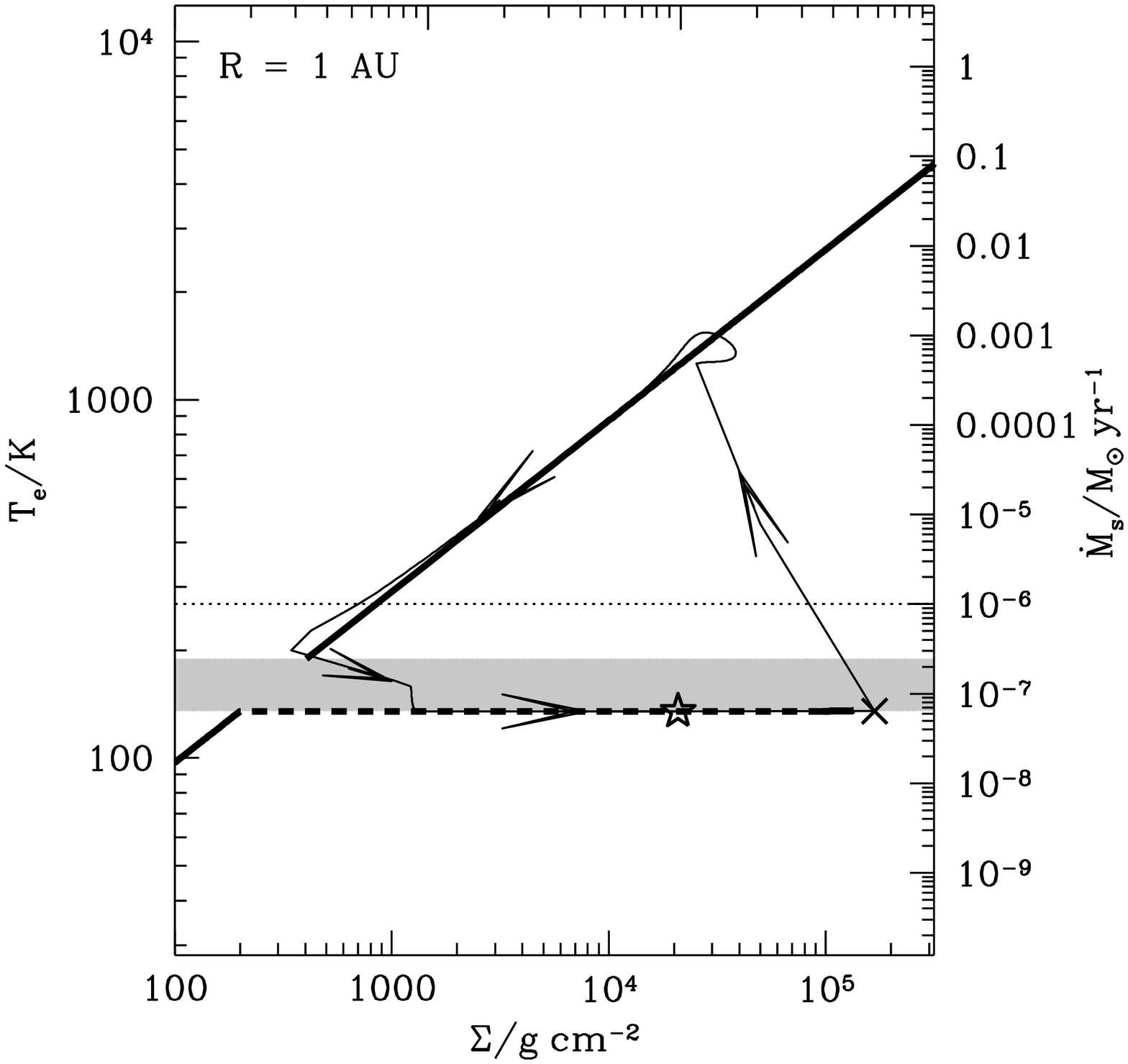}
\includegraphics[width=7cm]{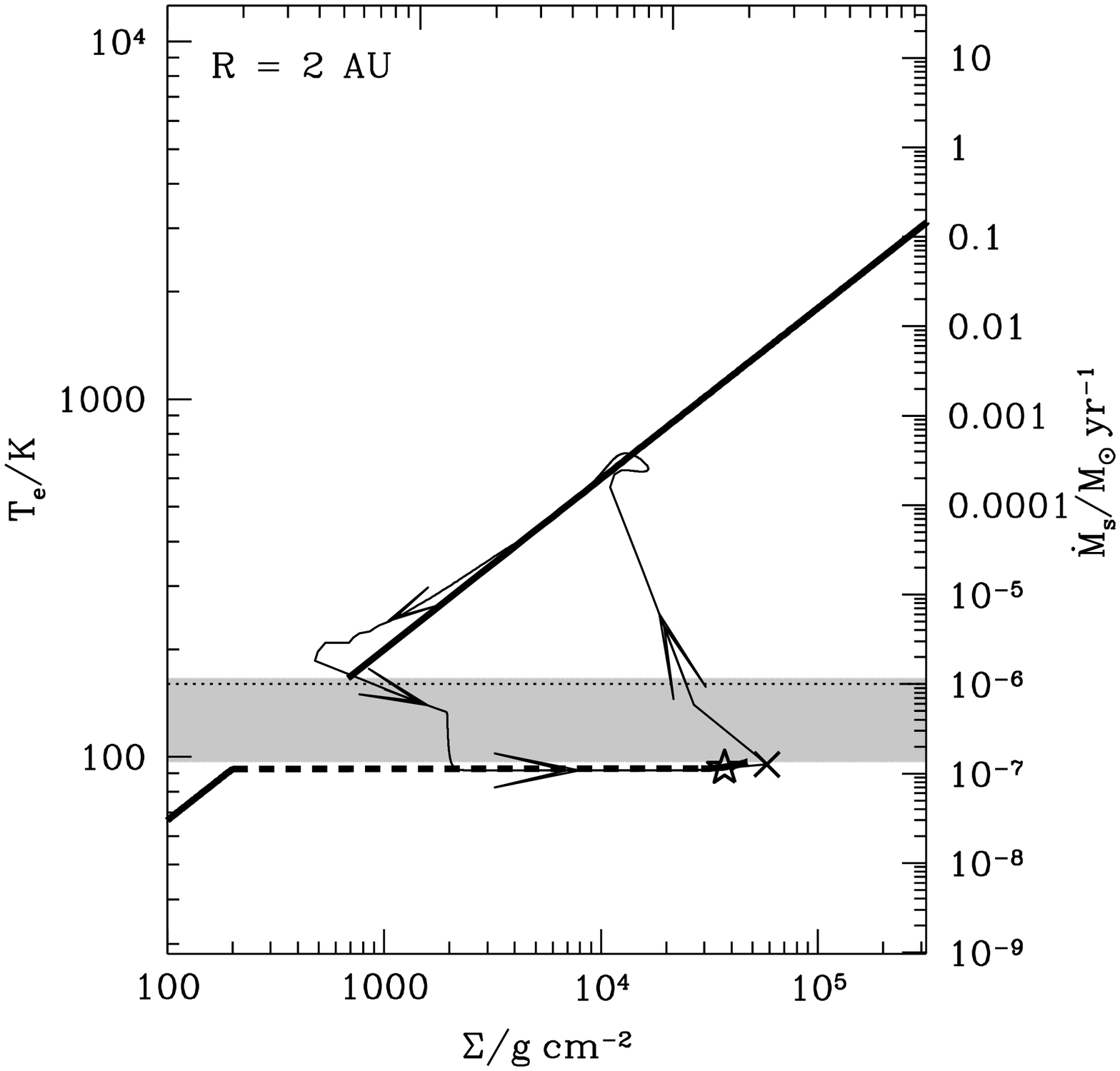}
\includegraphics[width=7cm]{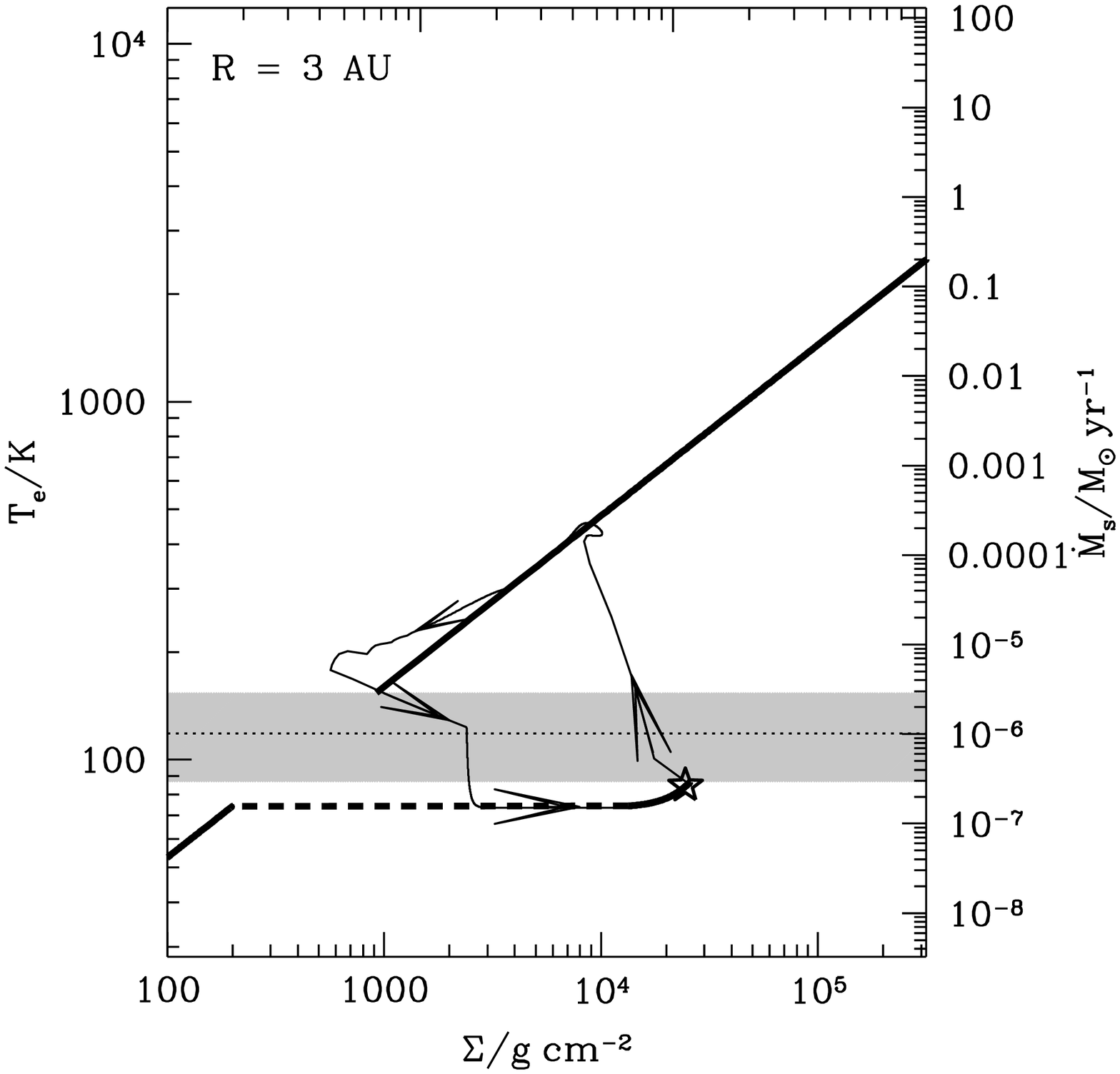}
\includegraphics[width=7cm]{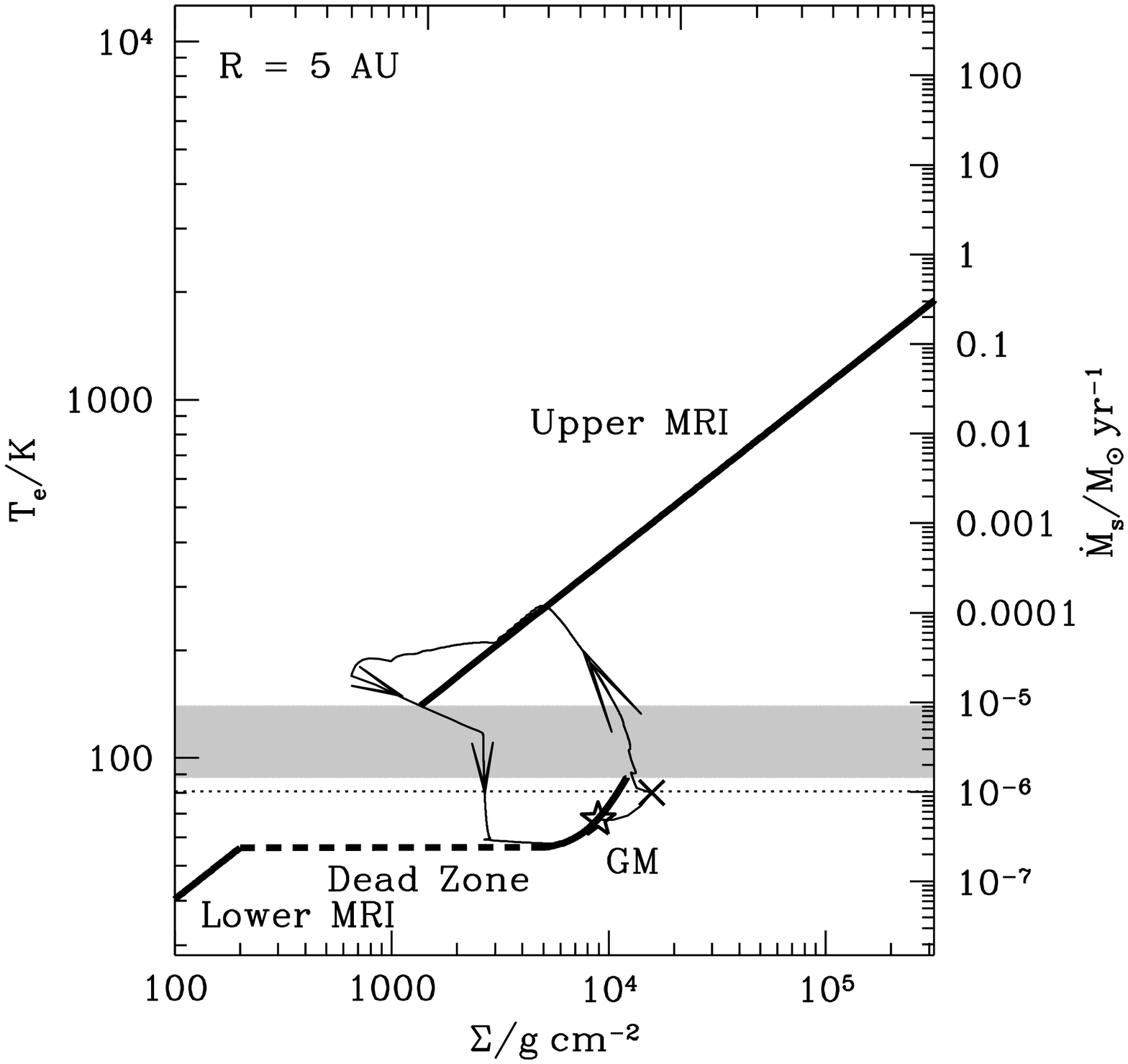}
\includegraphics[width=7cm]{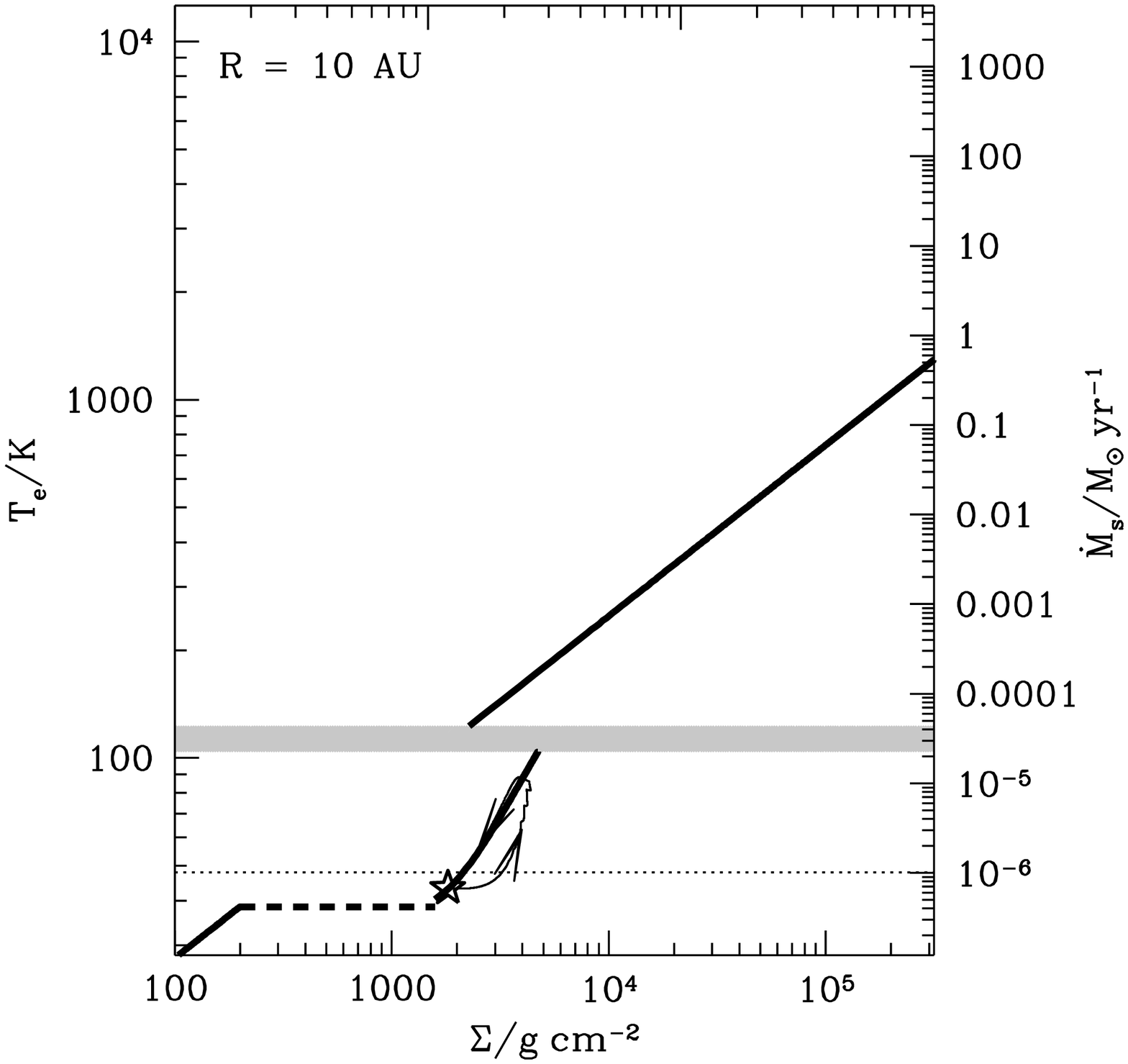}
\caption{The state diagrams at radii $R=0.1$, $1$, $2$, 3, 5 and
  $10\,\rm au$. The thick solid lines show the analytical steady
  states. The thick dashed line shows the dead zone branch. The thin
  lines with arrows show the global time-dependent numerical
  models. The dotted line shows the infall accretion rate that is
  equal to the steady state accretion rate through the disc. The
  shaded region is where there is no steady state solution and the
  disc is locally unstable. The star shows where the outburst is
  triggered globally and the cross where it is triggered locally. }
\label{data}
\end{figure*}

In this work we take an example of a circumstellar disc, but note that
the same mechanisms apply to circumplanetary discs on a smaller
scale. We choose a star of mass $M=1\,\rm M_\odot$ and a disc with a
\cite{shakura73} viscosity parameter $\alpha_{\rm m}=0.01$, $T_{\rm
  crit}=800\,\rm K$ and $\Sigma_{\rm crit}=200\,\rm
g\,cm^{-2}$. Fig.~\ref{data} shows the state diagrams of the steady
solutions at various radii in the disc. The surface temperature is
shown against the surface density for the steady state solutions at a
fixed radius. The surface temperature in the disc is related to an
effective, steady accretion rate, $\dot M_{\rm s}$, with
\begin{equation}
T_e =\left(\frac{3\dot M_{\rm s}\Omega^2}{8 \pi \sigma}\right)^\frac{1}{4}.
\end{equation}

All radii shown, except $R=0.1\,\rm au$, show the three steady
branches. There are two fully MRI active solutions, the upper branch
with $T>T_{\rm crit}$ and the lower branch with $\Sigma<\Sigma_{\rm
  crit}$. The third branch is the self gravitating GM solution. The
steady state solutions are shown in the solid thick lines and the dead
zone branch (that is not a steady branch) is shown as the dashed
line. The branches are labeled in the state diagram at $R=5\,\rm au$.
The shaded region is where there is no steady state solution and the
disc is locally unstable.  If the imposed disc accretion rate (infall
accretion rate onto the disc), represented as the dotted line, lies
within the shaded region, then the disc is locally unstable at that
radius.  The state diagram at a radius of $R=0.1\,\rm au$ does not
have an unstable region because a fully MRI active disc solution
exists at all accretion rates because either $\Sigma<\Sigma_{\rm
  crit}$ or $T>T_{\rm crit}$.

\subsection{Innermost Unstable Radius}

For a disc with an unstable region, the innermost radius that is
locally unstable occurs where the thermally ionised fully MRI active
solution has a mid-plane temperature equal to the critical
temperature. The surface density is larger than the critical that is
ionised by external sources, if the unstable region exists.

This location can been seen in the state diagrams to occur at a radius
where the accretion rate through the disc is such that it lies on the
lower end of the upper MRI branch (similar to the middle left plot in
Fig.~\ref{data}).  That is, the radius where the dotted line
intersects the lower end of the upper, solid sloped line.  The upper
MRI branch begins where the mid-plane temperature is equal to the
critical, $T=T_{\rm crit}$. We scale the variables to $\dot M'=\dot M
/(10^{-6}\,\rm M_\odot\, yr^{-1})$, $M'=M/{\rm M_\odot}$,
$\alpha'=\alpha/0.01$ and $T_{\rm crit}'=T_{\rm crit}/{800\,\rm
  K}$. We find steady state fully turbulent solutions (with
$\Sigma_{\rm m}=\Sigma$ and $\nu_{\rm g}=0$). The radius at which the
fully turbulent solution has the critical mid-plane temperature is
\begin{equation}
R= 1.87 \,  \frac{ \dot M'^{4/9} M'^{1/3}}{\alpha'^{2/9} T_{\rm crit}'^{14/15}}
\,\rm au.
\end{equation}
This is the innermost radius that is locally unstable to the GM
instability.

\subsection{Outermost Unstable Radius}

The outermost radius for which no steady solution exists (and thus the
disc is locally unstable to the GM instability) occurs where the GM
solution has a mid-plane temperature equal to the critical. In the
state diagram this is where the accretion rate through the disc lies
at the upper end of the GM branch (similar to the lower left plot in
Fig.~\ref{data}). This is where the self gravitating, GM solution
reaches the critical mid-plane temperature. We again solve the steady
state equations, but now include the self gravitating term. However,
the equations must be solved numerically. For an accretion rate of
$\dot M_{\rm infall}=10^{-6}\,\rm M_\odot\, yr^{-1}$, we find the
radius to be $R=4.51\,\rm au$. Thus, the locally unstable region for
this disc extends from $R=1.87\,\rm au$ to $R=4.51\,\rm au$. Outside
of this region there is a steady GM disc solution, and inside there is
a fully MRI solution. We note that the innermost and outmost unstable
radii in the disc do not necessarily correspond to the inner and outer
edges of a dead zone in a time-dependent disc.

\subsection{Outer Transition to MRI Branch}

\begin{figure*}
\includegraphics[width=7.5cm]{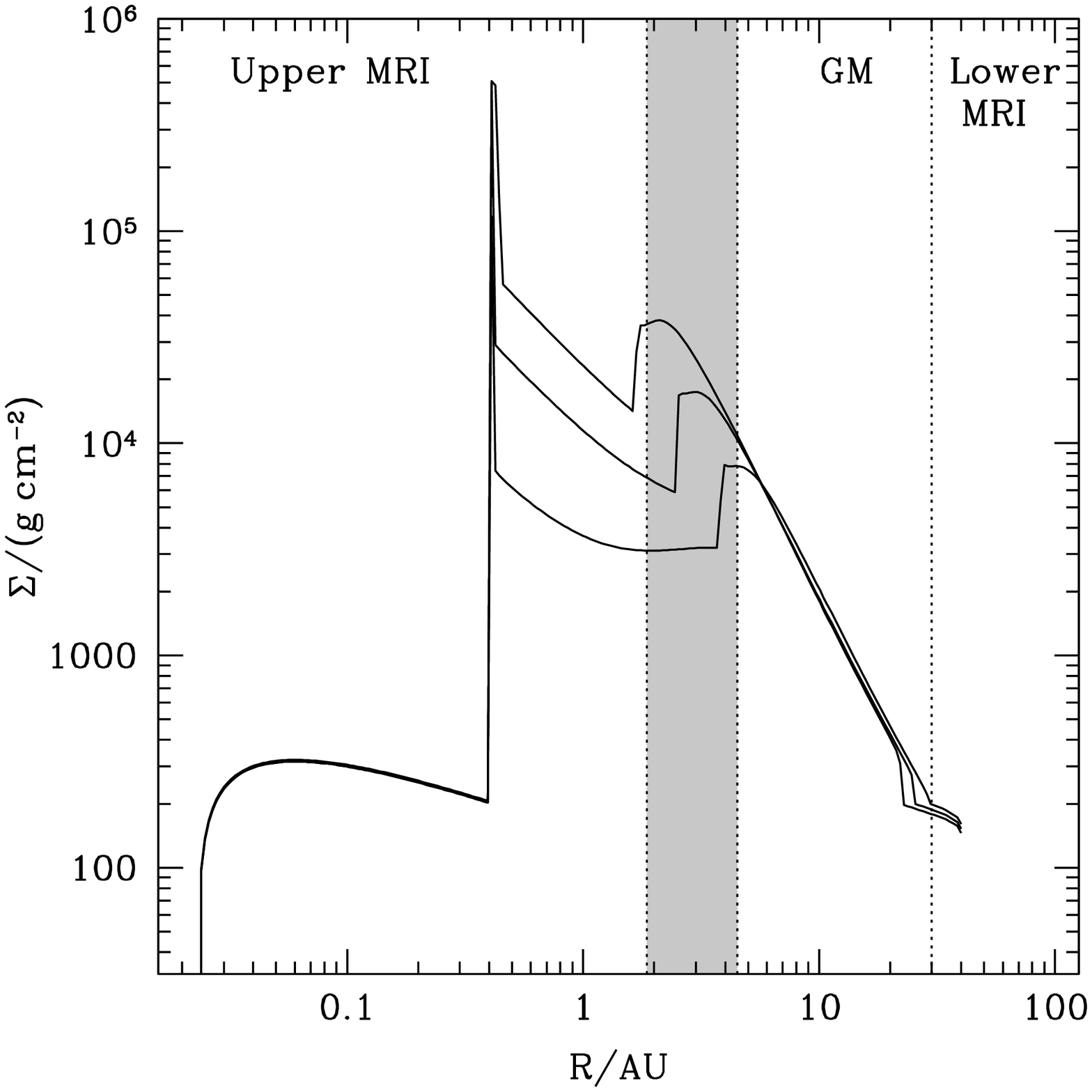}
\includegraphics[width=7.5cm]{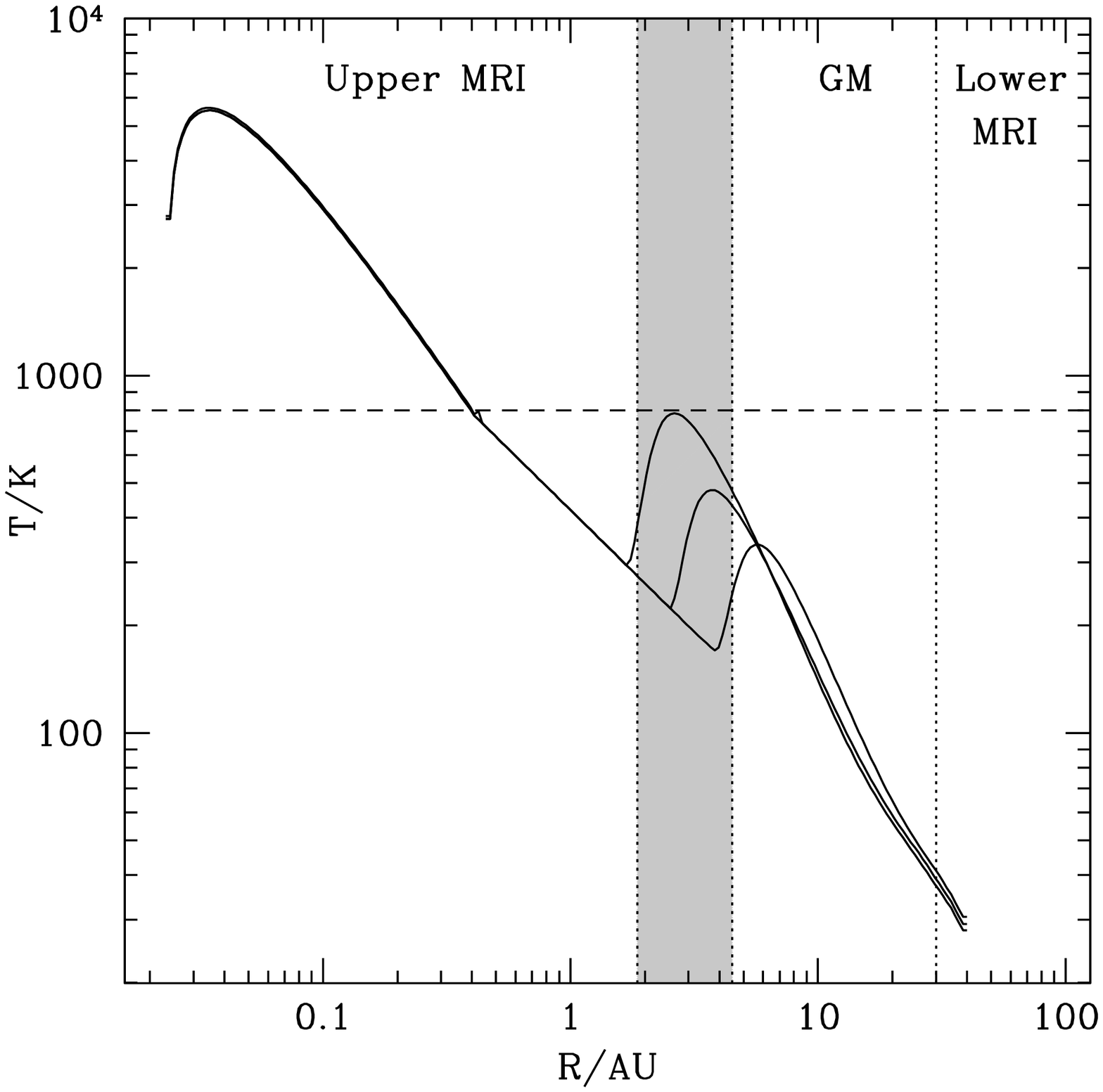}
\caption{Disc structure leading up to an outburst. Left: The total
  surface density. Right: The mid-plane temperature. The shaded
  regions show the unstable regions where no steady state disc
  solution exists. The regions are labelled where steady state, MRI
  and GM solutions exist. The profiles, in order of increasing height
  within the unstable region, are at times $t=1.22\times 10^4\,\rm
  yr$, $5.21\times 10^4\,\rm yr$ and $1.12\times 10^5\,\rm yr$ since
  the previous outburst. The final time shows the disc just before the
  outburst is triggered. The dashed line in the temperature plot shows
  the critical temperature, $T_{\rm crit}$, above which the MRI drives
  turbulence.}
\label{st}
\end{figure*}

The outer parts of the disc are fully MRI active where the surface
density is smaller than the critical, $\Sigma<\Sigma_{\rm crit}$. This
occurs at a radius
\begin{equation}
R=29.15\, \frac{\dot M'^{11/9} M'^{1/3}}{\alpha'^{16/9}} \,\rm au.
\end{equation}
Hence, in summary, in the region $R<1.87\,\rm au$ the disc has a fully
MRI active solution, $1.87<R/{\rm au}<4.51$ the disc has no steady
solution (and is locally unstable to the GM instability). In
$4.51<R/{\rm au}<29.2$ there is a steady GM solution. In $R>29.2\,\rm
au$ the disc is fully MRI active. In Figs.~\ref{st} and~3 we label
these regions as a function of radius in the disc and shade the region
for which there is no steady solution.

\section{Global Disc Instability}

\begin{figure*}
\includegraphics[width=7cm]{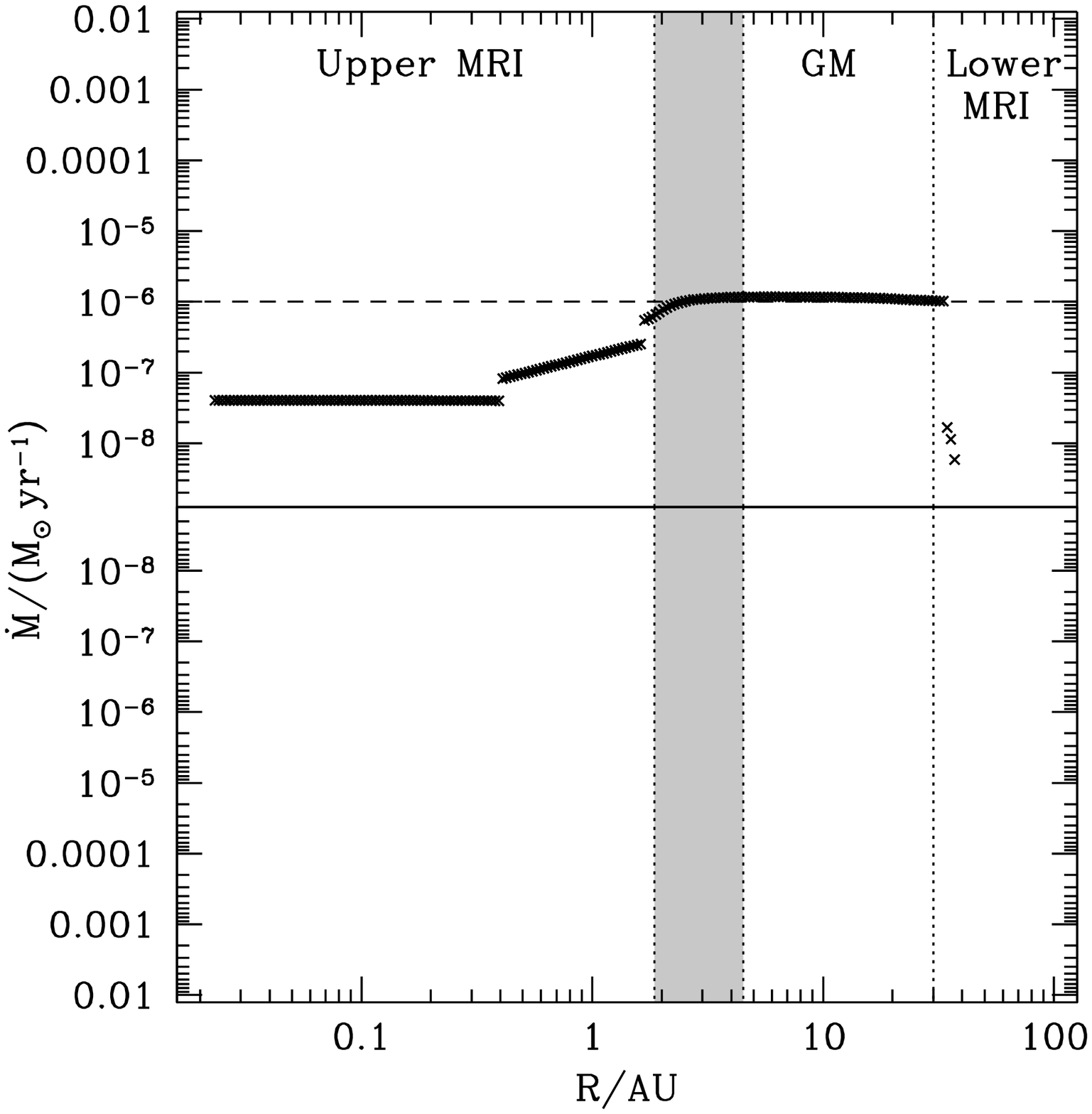}
\includegraphics[width=7cm]{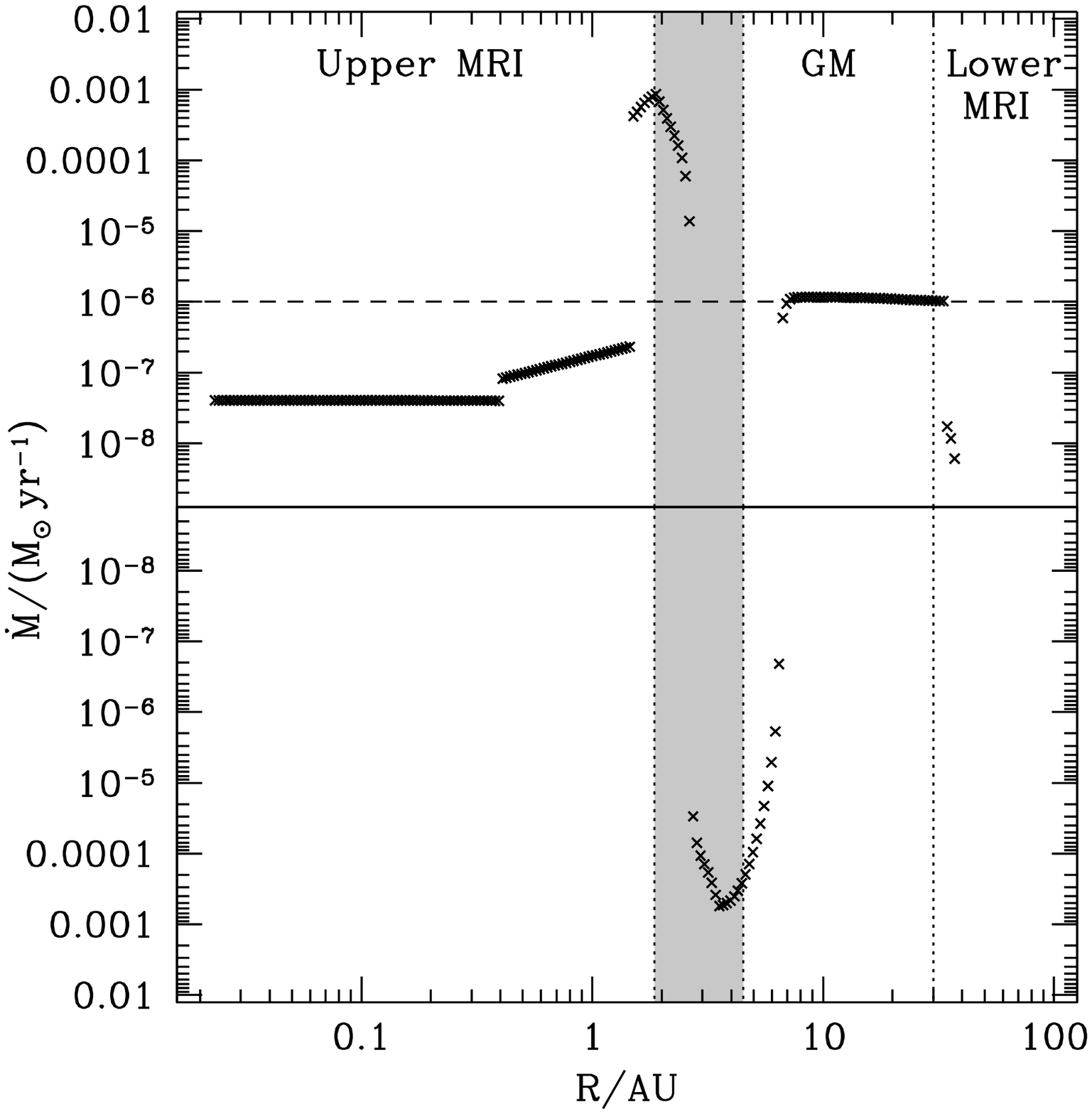}  
\includegraphics[width=7cm]{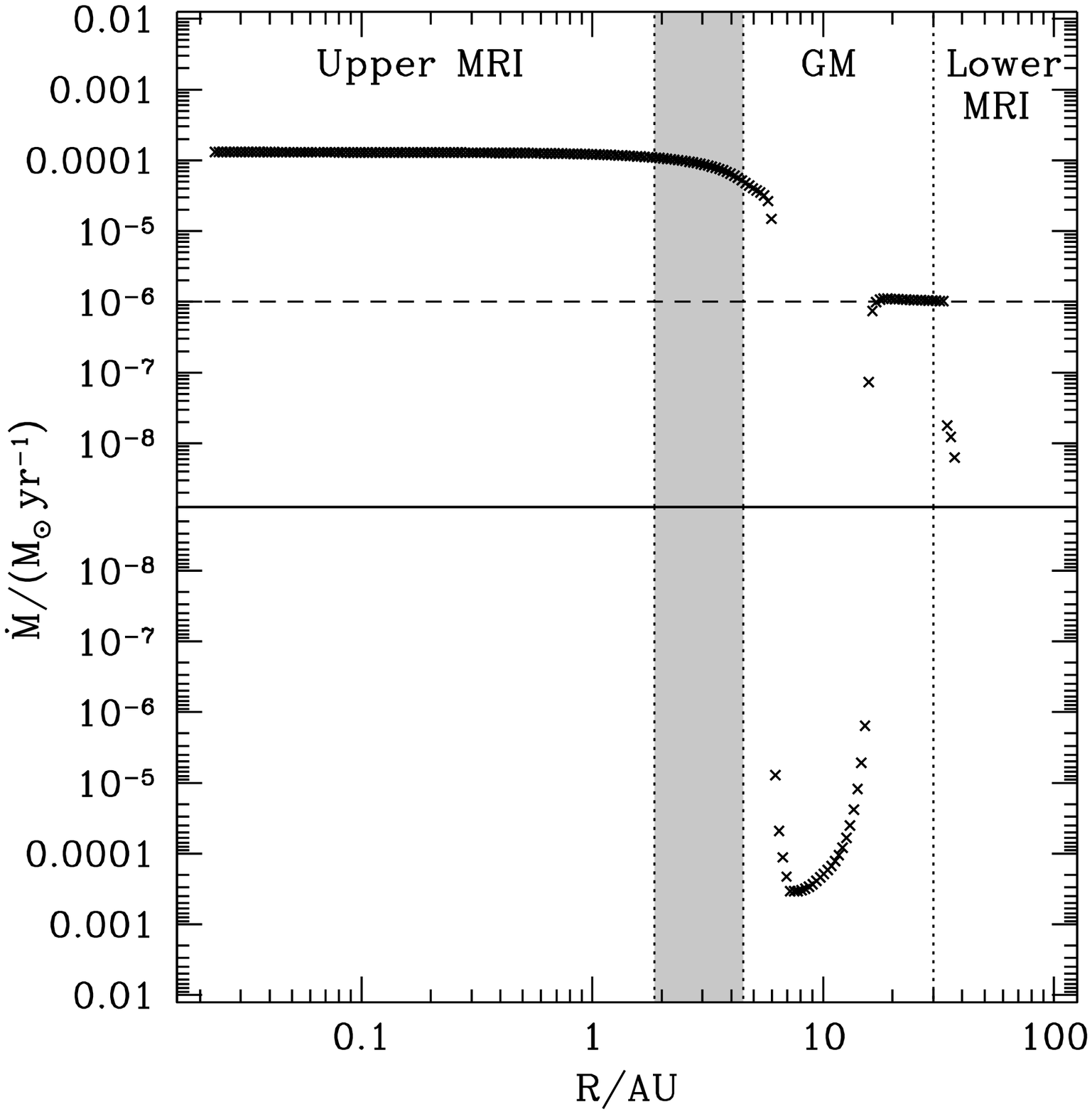}
\includegraphics[width=7cm]{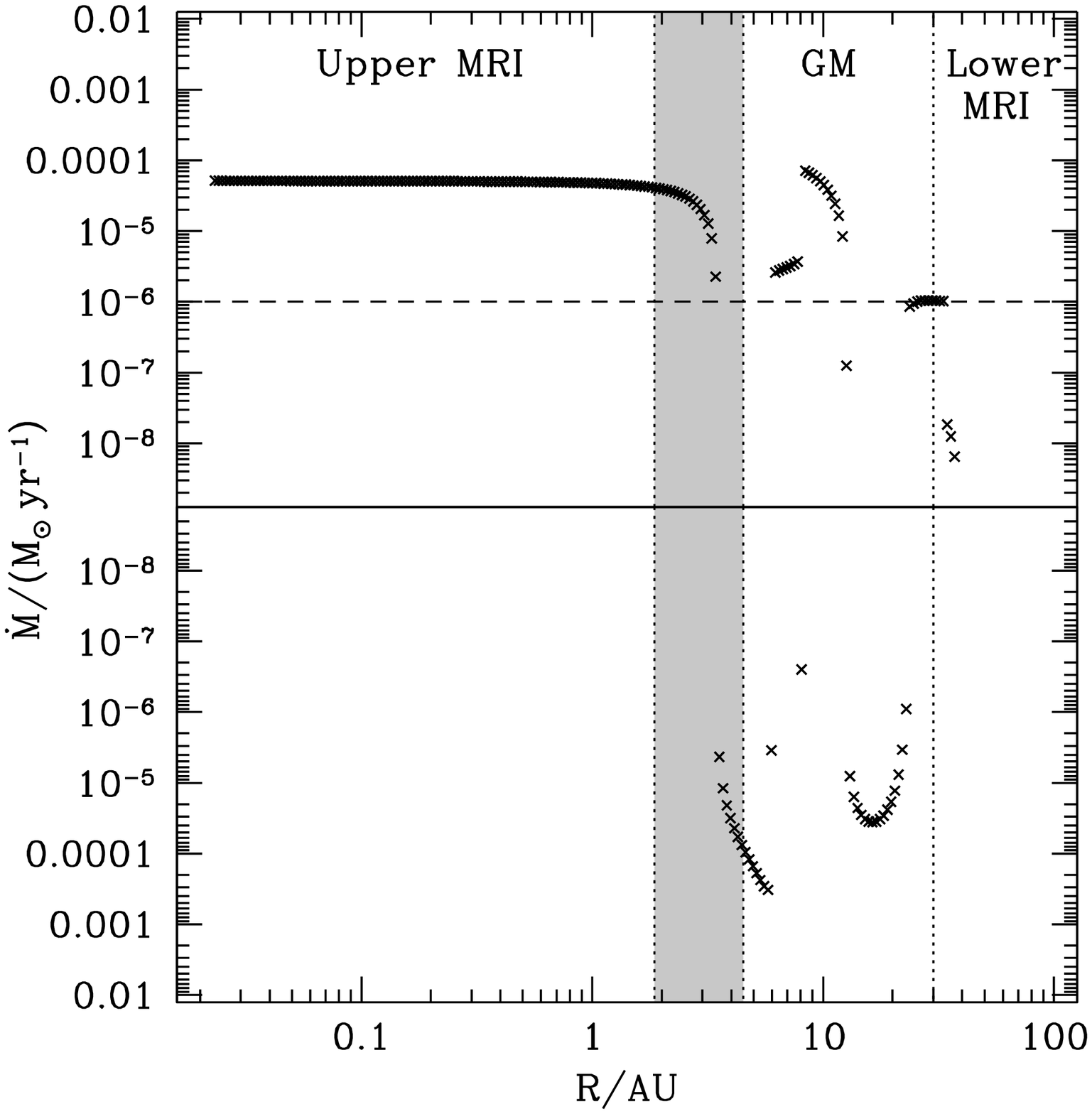}
\includegraphics[width=7cm]{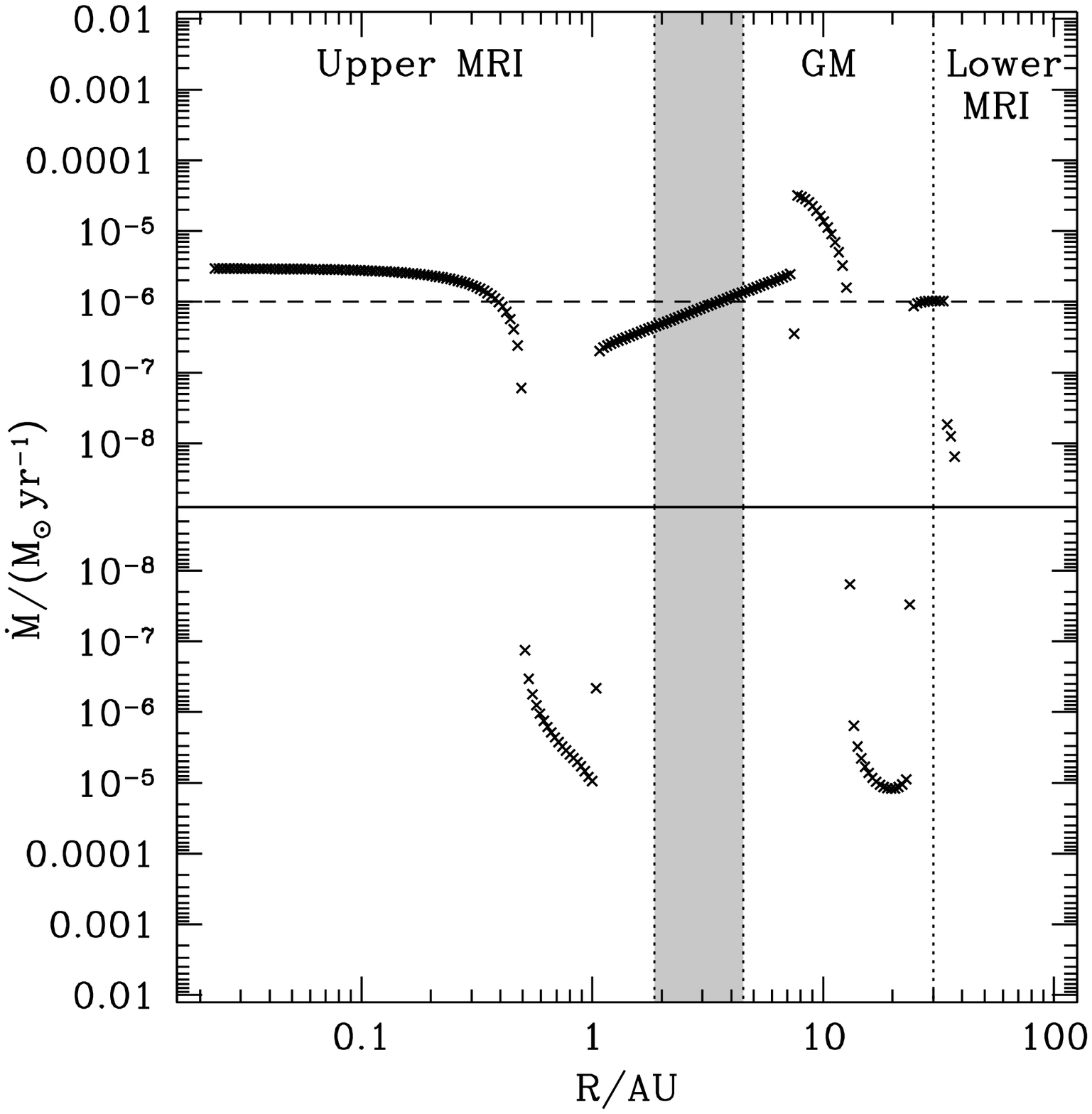}
\includegraphics[width=7cm]{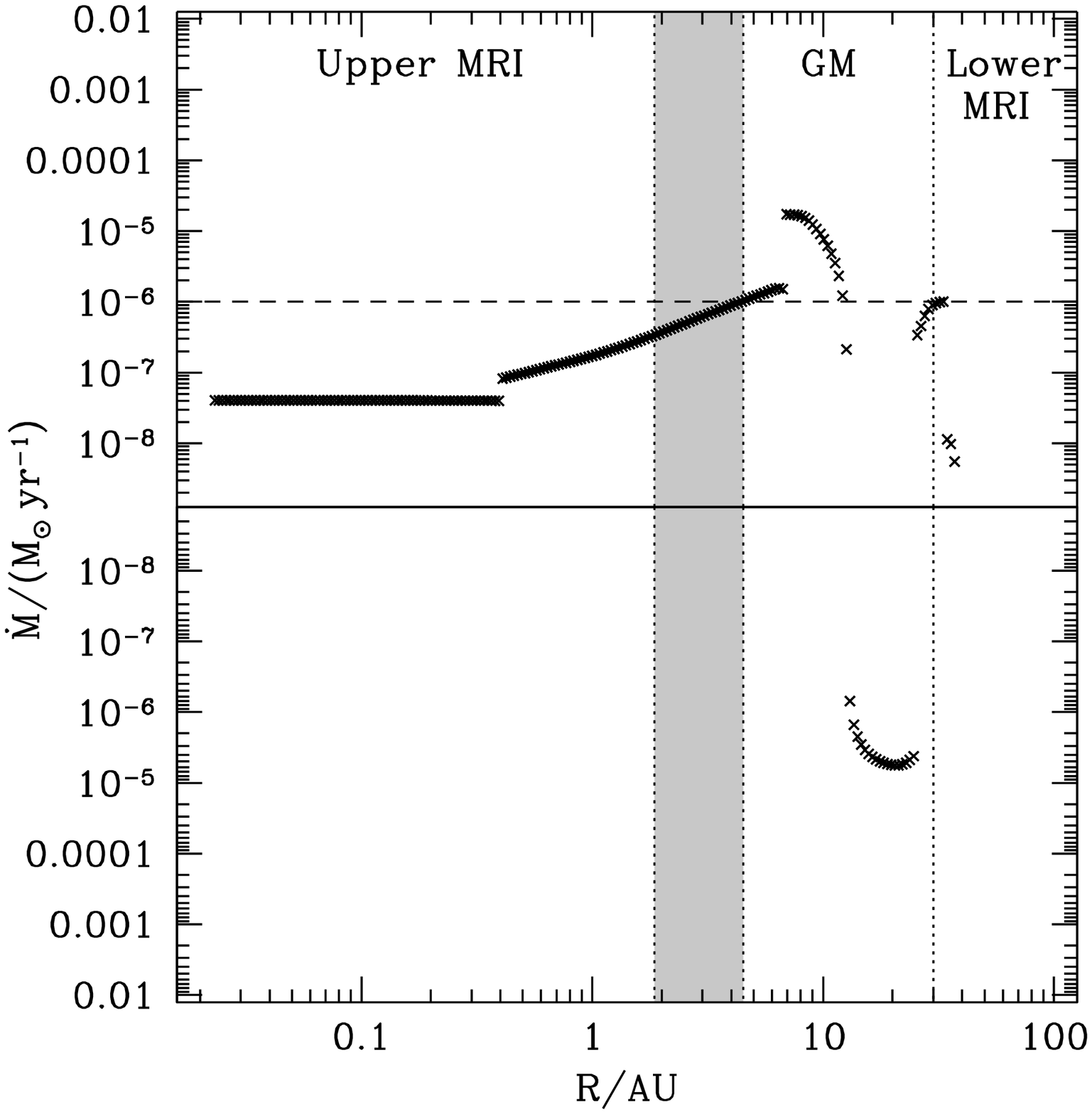}
\caption{The radius dependent accretion rate through the disc during
  the outburst cycle. The plots shown are at times $t=0$, $140$,
  $400$, $800$, $1000$ and $1400\,\rm yr$ from the start of the
  outburst. In each plot, the upper portion shows the inward accretion
  and the lower portion shows the outward accretion.}
\label{mr}
\end{figure*}

We solve the time-dependent accretion disc equations that consist of a
surface density evolution equation \citep{pringle81} and an energy
equation \citep{pringle86,cannizzo93}.  The equations are described in
more detail in \cite{martin11} (see their equations~1 and~3). We take
a grid of 200 points equally spaced in $\log R$ that extends from
$R_{\rm in}=5\,\rm R_\odot$ to $R_{\rm out}=40\,\rm au$
\citep[e.g.][]{armitage01,martin07}.  At the inner boundary we have a
zero torque boundary condition to allow the inward flow of gas on to
the star. At the outer boundary, we choose a zero radial velocity
condition to prevent material from leaving the disc there. There is a
constant infall accretion rate on to the disc of $\dot M_{\rm
  infall}=10^{-6}\,\rm M_\odot\,yr^{-1}$ at a radius of $35\,\rm
au$. As shown in the previous section, for this accretion rate the
disc is locally unstable for a range of radii.

The disc evolves through several outbursts before the limit cycle
repeats. Thus, the initial conditions of the disc have no effect on
the limit cycles shown.  Fig.~\ref{data} shows the local state
diagrams, for various radii, of the surface temperature against the
surface density. The thick lines show the analytic steady state
solutions (as described in the previous section) and the thin lines
show the numerical time-dependent evolution. The dashed line shows the
steady infall accretion rate.

The crosses mark the position at the time when the outburst is
triggered at that radius (i.e. where $T>T_{\rm crit}$), and the stars
mark the position when the outburst is first triggered globally. For
the $R=3\,\rm au$ case, the two positions coincide with the end of the
GM branch and hence also with a temperature of $T=T_{\rm
  crit}$. However, at other radii, the position when the outburst is
triggered locally does not coincide with the position where it is
triggered globally. Once the outburst is triggered somewhere in the
disc, it propagates very quickly and adjacent radii undergo their
limit cycles similarly fast. At $R=0.1\,\rm au$, the mid-plane
temperature of the disc does not drop below the critical during the
cycle, hence there is no cross shown. Similarly, at $R=10\,\rm au$,
the mid-plane temperature during the cycle does not exceed the
critical temperature and so there is no cross shown.

The outburst propagates inwards and hence all radii inside of the
trigger radius show outburst behaviour. The total mass of the disc
before the outburst is $0.45\,\rm M_\odot$ of which $0.34\,\rm
M_\odot$ is within the dead zone and $0.024\,\rm M_\odot$ is not
self-gravitating. During the outburst, a mass of $0.1\,\rm M_\odot$ is
accreted on to the star. This amount of material is initially
distributed up to a radius of $2.9\,\rm au$ (that roughly coincides
with the trigger radius).  The outburst also propagates outwards. The
dead zone has a high surface density that suddenly becomes more
turbulent when the MRI is triggered, causing material to move
outwards. For example, at $R=5\,\rm au$, the disc has a steady state
GM solution. However, this solution is not found because of the
outward propagation. The disc is depleted after an outburst, leaving a
less massive and cooler disc than the steady state would predict.  The
disc moves towards the steady state but the outburst is first
triggered at an accretion rate lower than the steady rate.  Even
further out, where the steady solution is lower down the GM branch,
the outward propagation is not sufficient to trigger an outburst and
the disc remains close to the GM branch and then moves downwards after
the outburst. The outermost radius for which the disc moves up to the
upper MRI branch during the outburst is around $R=9\,\rm au$, twice
the radius of outer edge of the locally GM unstable region.

Fig.~\ref{st} shows the disc structure leading up to an outburst.  The
branch labels in the figure (Upper MRI etc.) refer to where steady
state solutions exist. For the time-dependent evolution being plotted,
these labels may not reflect the instantaneous structure at each
radius.  The shaded region shows where there is no steady state disc
solution.  The innermost parts of the disc are fully MRI
turbulent. However, beyond radius of $R=0.41\,\rm au$, the mid-plane
temperature of the disc drops below the critical value. At this
radius, $R=0.41\,\rm au$, there is a sharp increase in the surface
density and this marks the inner edge of the dead zone. The narrow
density peak is due to a local pileup of gas.  It is the result of the
outward spread of turbulent gas from the active, inner (inside 0.41
AU) region into the nonturbulent dead zone where it cannot spread
outward any further.  The narrowness is a consequence of the sharp
transition criteria we apply between active and dead zones. If we
applied a smoother transition criteria, the peak would be spread out
over a larger range of radii. The dead zone covers some of the region
where there exists a steady state MRI disc solution (as labeled Upper
MRI). This overlap exists because the outer parts of the flow near
$R=1\,\rm au$ are limited to the accretion rate of the active surface
layer. There is not sufficient flow in this region to reach the steady
solution at the infall accretion rate.

The local peak in the surface density within the unstable region marks
the radius in the disc at which the dead zone becomes self
gravitating. This local peak moves inwards in time as material builds
up. There is a similar local peak in the temperature profile that
moves inwards slightly behind the surface density peak in time. Once
this peak reaches the critical temperature required for the MRI, the
outburst is triggered (as in the last time shown in
Fig.~\ref{st}). The outburst is first triggered at a radius of
$R=2.6\,\rm au$.

It is the outer parts of the dead zone that become self gravitating
first (as shown in Fig.~\ref{st}). Hence, in the limit cycles in
Fig.~\ref{data}, the larger the radius, the sooner the disc reaches
the GM branch. However, the speed at which the disc moves up this
branch, decreases with radius. As a function of radius, there is then
opposite behavior in the time that the disc reaches this branch and
the speed at which it travels up the branch. Hence, the outburst is
triggered somewhere in the middle of the locally unstable region.

Fig.~\ref{mr} shows the radius dependent accretion rate through the
disc at various times during the outburst cycle. The upper portion of
each plot represents the inward flow and the lower portion the outward
flow. The top left plot shows the accretion rate just before an
outburst, that corresponds to the upper lines for the surface density
and temperature shown in Fig.~\ref{st}. There are three main parts to
the disc, the inner part that is fully MRI active, the middle part
where there is a dead zone and the large outer part that is self
gravitating. In the very outermost parts of the disc, the accretion
drops off because there is a zero flow outer boundary condition. This
only affects the parts of the disc outside of the radius where
material is injected, at $35\,\rm au$.

The top right hand plot shows the accretion rate shortly after the
outburst is triggered. There are both inward and outward propagating
waves moving away from the trigger point at $2.6\,\rm au$. This is the
snow plough effect, material moves both inwards and outwards away from
the trigger radius. It is this snow plough that causes the various
radii to be coordinated. The middle left plot shows the inner wave has
propagated all the way in and the disc is now in outburst phase with a
high accretion rate on to the star. The outward propagating waves
continue to move outwards. In the middle right hand plot, the dead
zone has begun to re-form. This initially occurs at a radius of
$8.1\,\rm au$. The dead zone formation propagates both inwards and
outwards as seen in the lower left plot.  As the dead zone forms, the
viscous torque there drops. Thus, material just interior to the dead
zone initially moves outwards because the outward viscous torque is
larger than the inward torque.  This is the cause of the narrow peak
in the surface density at the inner edge of the dead zone (as
described in Fig.~2). Eventually the inner parts of the disc move back
to the quiescent phase as the dead zone blocks the flow, as seen in
the bottom right hand plot. This marks the end of the outburst,
although there is still some ongoing outward propagation in the outer
regions. The material in the disc builds up again as shown in
Fig.~\ref{st} until the cycle repeats.

\section{Discussion}

There are a number of simplifications in our model that should be
investigated in future work. Some of these were discussed in more
detail in \cite{martin11} and \cite{lubow12a}.  For example, we have
assumed that the critical surface density that is ionised by external
sources in constant with radius. However, a more realistic way to find
the extent of the dead zone may be with a critical magnetic Reynolds
number \citep[e.g.][]{fleming00, matsumura03}. The critical value
however remains uncertain \citep[e.g.][]{martin12b}. The ionisation
may be further suppressed by effects such as ambipolar diffusion and
the presence of dust and polycyclic aromatic hydrocarbons
\citep[e.g.][]{bai09,bai11,perez11,simon12,dzyurkevich13}. No matter
how the dead zone is determined, the outburst mechanism works in the
same way. The shape of the limit cycle may be slightly different
\citep[see][]{martin12a}, but the principles remain the same.

There remains some uncertainty in the prescriptions for the
viscosities in the disc.  Observations of the fully MRI turbulent
discs in dwarf novae and X-ray binaries suggest $\alpha \approx
0,1-0.4$ \citep{king07}. However, in discs around T Tauri stars, such
a high $\alpha$ would cause the outer discs to expand too quickly
\citep{hartmann98}. Moreover, theoretical work suggests a smaller
$\alpha \approx 0.01$ \citep[e.g.][]{fromang07}. We have taken
$\alpha_{\rm m}=0.01$ in this work, but note that the qualitative
evolution of the disc would be similar had we taken a larger value for
$\alpha_{\rm m}$. There is also some uncertainty in the value of
$\alpha_{\rm g}$ in a self gravitating disc. But its exact value has little influence
of the disc structure and evolution, as noted by \cite{zhu10a}.
Within the self
gravitating region the Toomre parameter in our model is nearly constant,
$Q\approx Q_{\rm crit}$. In our prescription the value of $\alpha_{\rm
  g}$ is not fixed and depends on the Toomre parameter, $Q$. In steady
state, implicitly it varies with the local cooling timescale, $t_{\rm
  cool}$, such that $\alpha_{\rm g}\propto 1/(t_{\rm cool}\Omega)$
\citep[see also][]{gammie01}.

In a real disc the infall accretion rate is not likely to be constant
in time, but rather decrease.  The locally unstable region would
decrease in time until the accretion rate is so low that a fully MRI
active solution is found. For the parameters chosen in this work, this
would be for accretion rates $\dot M_{\rm infall} \lesssim
10^{-8}\,\rm M_\odot\, yr^{-1}$.

The existence of dead zones in protoplanetary discs may have important
implications for planet formation.  Planets that survive after the
disc has dispersed \citep[e.g. by
  photoevaportation][]{hollenbach94,alexander06} likely formed after
the final GM disc outburst. If planets form between outbursts, they
likely accrete on to the star during the outburst. As the infall
accretion rate drops over time, the outbursts cease but a dead zone may
remain. This dead zone provides a quiescent region suitable for planet
formation. The presence of a dead zone also affects the migration of
planets through the disc. Planets embedded in low viscosity discs
can undergo very slow and sometimes chaotic migration \citep{li09,yu10}.
The nature of gap opening and migration of
planets through a layered disc should be investigated in future work.

\section{Conclusions}

The GM disc instability can be explained as transitions between steady
state solutions in a state diagram that plots the surface temperature
against the surface density at a fixed radius.  We have examined local
cycles at different radii. There is a range of radii for which a given
disc is locally GM unstable. For the parameters chosen in this work
($\Sigma_{\rm crit}=200\,\rm g\,cm^{-2}$, $T_{\rm crit}=800\,\rm K$,
$\dot M_{\rm infall}=10^{-6}\,\rm M_\odot\, yr^{-1}$ and $\alpha=0.01$), we find
the unstable range to be $1.87<R/{\rm au}<4.51$.

The outburst is first triggered in the middle of the range of locally
unstable radii. For the parameters we chose, the trigger radius is at
$2.6\,\rm au$. The resulting snow plough of material that propagates
both inwards and outwards links the different radii in the disc. All
radii inside of the trigger radius become unstable and the mass in
this region moves inwards to be accreted during the outburst.  The
snow plough effect pushes material outwards through the disc to radii
which would otherwise lie along the steady GM branch and be locally
stable. The higher density causes these outer radii to also become
unstable. The outburst propagates to a radius of around $9\,\rm au$,
about double that of the outermost locally unstable radius. Even
further out than this, there are increases in density but these are
not sufficient for a change in the state.

\section*{Acknowledgements}

We thank the anonymous referee for useful comments. RGM's support was
provided in part under contract with the California Institute of
Technology (Caltech) funded by NASA through the Sagan Fellowship
Program. SHL acknowledges support from NASA grant NNX11AK61G.


\label{lastpage}
\end{document}